# Algorithms are not neutral: Bias in collaborative filtering

Catherine Stinson

Philosophy Department and School of Computing
Queen's University
Kingston, ON Canada
c.stinson@queensu.ca

Abstract

Discussions of algorithmic bias tend to focus on examples where either the data or the people building the algorithms are biased. This gives the impression that clean data and good intentions could eliminate bias. The neutrality of the algorithms themselves is defended by prominent Artificial Intelligence researchers. However, algorithms are not neutral. In addition to biased data and biased algorithm makers, AI algorithms themselves can be biased. This is illustrated with the example of collaborative filtering, which is known to suffer from popularity, and homogenizing biases. Iterative information filtering algorithms in general create a selection bias in the course of learning from user responses to documents that the algorithm recommended. These are not merely biases in the statistical sense; these statistical biases can cause discriminatory outcomes. Data points on the margins of distributions of human data tend to correspond to marginalized people. Popularity and homogenizing biases have the effect of further marginalizing the already marginal. This source of bias warrants serious attention given the ubiquity of algorithmic decision-making.

## Introduction

There is growing awareness that the outcomes of algorithms can be discriminatory. The best known recent examples are ones where the data used to train machine learning (ML) algorithms are systematically biased, leading to algorithms with discriminatory outcomes. Cases have been uncovered where using data about past decisions to train systems to make policing, hiring, or credit decisions means that historical discrimination gets programmed into the algorithm, perpetuating the bias in future decisions (Angwin et al. 2016).

Biased datasets can be the downstream result of developer teams lacking diversity. That facial recognition algorithms are an order of magnitude less accurate for black female faces than for white male faces has been attributed to the lack of black and female faces among the training examples used to build facial recognition systems. That lack of diversity in the training examples in turn stems from a lack of gender and racial diversity among AI researchers (Buolamwini and Gebru 2018).

In these cases of biased datasets, 'bias' can refer to (at least) two distinct things. In statistics and ML, 'selection bias' refers to a non-random process being used to select a sample from a population. An example is if people conducting a marketing survey only ask black men to participate. The resulting dataset would then over-represent the opinions of black men. If the opinions of black men happen to be different on average than those of the general population, the survey results would be misleading.

The colloquial meaning of bias is closer to the definition Friedman and Nissenbaum offer of "bias of moral import," which is, "systematically and unfairly discriminat[ing] against certain individuals or groups of individuals in favor of others" (Friedman and Nissenbaum 1996). If a company fires a group of LGBTQ employees without cause, that would be this second kind of bias or discrimination.

Another well-known source of algorithmic bias is the people building algorithms. There are documented cases where algorithms have been designed specifically to create discriminatory outcomes. Redlining certain neighbourhoods as high risks for mortgages, based on the racial composition of residents, or choosing to target only men to show certain kinds of job ads (Dwoskin 2018) are two examples. In most cases, bias is accidental and unforeseen, resulting from the limited perspective of algorithm makers. Products might be built for the benefit of one group, while inadvertently producing negative side-effects for others, such as speech recognition algorithms failing to work for users with non-standard accents, or YouTube's click-maximizing algorithms benefitting advertizers at the expense of website users (Tufekci 2018).

The choice of research questions to pursue, or applications to develop can also overlook the needs of some groups, such as health apps that do not include period trackers. Testing to ensure safety, usefulness and performance can likewise fail to consider the needs of some groups, like automatic soap dispensers that do not reliably detect dark hands.

Here the focus is on a different locus of bias: the algorithm itself. This source of bias is not addressed in high profile reviews of algorithmic bias (Friedman and Nissenbaum 1996; Barocas and Selbst 2016). That algorithms are neutral is a popular misconception, even among AI reserchers. On Twitter, Yann LeCun declared, "People are biased. Data is biased... But learning algorithms themselves are not biased" (LeCun 2019).

This paper challenges these claims by highlighting bias in collaborative filtering, a type of ML algorithm commonly used in recommender systems. Below I review the statistical biases inherent in collaborative filtering and iterative information filtering. I then spell out how statistical biases can translate into discrimination.

## Bias in Collaborative Filtering

Collaborative filtering algorithms are used in popular recommender systems, that show users items based on criteria like "Customers who viewed this item also viewed" or "Because you watched…" User profiles are constructed based on explicit ratings such as likes, and implicit ratings like viewing time. To come up with recommendations, the user profile is compared to other users' profiles to find matches. Items that were rated highly by users with similar profiles but that have not been seen, are then recommended to the user. User profiles are regularly updated as the user interacts with the system, with the goal of making the profile a more accurate predictor of the user's behaviour over time. Below I outline a few of the specific ways in which collaborative filtering algorithms are biased. Olteanu et al. (2019) catalogue a number of additional biases.

Selection Bias

For many popular recommender systems, ratings are sparse relative to the number of items available: most of the videos on Netflix have not been viewed or rated by most users. The algorithms used to predict user preferences are designed to have high prediction accuracy on the assumption that the missing ratings are missing at random, i.e., that there is no bias operating over which items are rated and which are not. This assumption that ratings are missing at random is false.

As the recommender narrows in on the user's tastes, it is simultaneously narrowing the scope of the data available to it on which to make those improvements. Most of the ratings the system uses, whether explicit or implicit, are for items that the user saw because the system recommended the item. The system cannot learn from the user's hypothetical ratings of things the user has not been shown. In order to do its job well, the algorithm would benefit from a broader base of ratings. In other words, collaborative filtering systems impose a selection bias on their own training data, then iteratively exacerbate that bias.

## Cold-Start Problem

The cold-start problem is a clear violation of the missing at random assumption. Although collaborative filtering was intended as a replacement for human reviewers, recommending new releases is a task they cannot perform. When a new item becomes available, there are no ratings of it by any user, meaning the recommended cannot recommend the item, unless a mechanism to counteract this bias is built in. In general, items in the system longer will build up more ratings over time, so be more likely to be recommended than newer items. This dynamic will develop even in a scenario where initial ratings are missing at random.

From the perspective of users, the cold-start problem appears as a (small c) conservative bias, where popular but older items are hard to avoid, and new things are harder to find. Likewise, the earlier an individual user gives a positive rating to an item, the more of an effect that item will have on their future recommendations, even if their tastes change or mature. In contrast, a more recent interest would have fewer total ratings associated with it, and thus exert less of an effect on recommendations.

## Popularity Bias

A closely related problem is known as popularity bias (Herlocker et al. 2004; Steck 2011), where very popular items are likely to get recommended to every user (and since recommendations make ratings more likely, popular items tend to increase in popularity). So even a user whose only positive ratings are for medieval Persian editions of ancient medical texts might get recommendations for *The Very Hungry Caterpillar*, simply because no matter what you like, it's likely that someone who liked the same has also bought *The Very Hungry Caterpillar*. Relatedly, a user might have bought *Fifty Shades of Gray* because they are writing a dissertation about representations of kink in popular culture, and end up having to wade through pulp romance novel recommendations that come highly rated by *Fifty Shades of Gray* fans, despite having no interest in the genre.

Profile injection attacks manipulate the probability of an item being recommended through the creation of fake user ratings. An infamous example is how the Amazon page for a book by anti-gay televangelist, Pat Robertson, listed an anal sex guide as a recommendation, after pranksters repeatedly viewed the two items together in order to form an association (Olsen 2002). This trick has also been used as a marketing ploy. Profile injection attacks illustrate the extent to which recommendations depend on popular patterns of ratings of other users.

The cold-start problem and popularity bias result from the number and timing of ratings not being evenly distributed among items in the dataset. These biases affect different users differently, and additional biases originate from the fact that users are not uniformly distributed in preference space. How much the user values novelty, how much the user's tastes have changed from their starting point, and how far their tastes lie from the mean can all vary. Many users' preferences will cluster around popular items, but other users will cluster in smaller niche groups (Horror fans, perhaps), and still others will have rare preferences (like our medieval Persian medical text fan), or atypical combinations of preferences (a fan of both Death Metal and musicals, for example).

## Over-specialization

Over-specialization occurs when a recommender algorithm offers choices that are much more narrow than the full range of what the user would like. In statistical terms this is not a problem of bias but of variance (the expectation of how far a variable deviates from its mean). Adamopoulos and Tuzhilin (2014) treat over-specialization as a problem stemming from an exclusive focus on prediction accuracy, while overlooking user-satisfaction.

Intuitively, the problem arises because items similar to those previously liked by a user will have a high probability of also being liked, even though what the user wants might be a wider range of recommendations that cover their preferences more fully. For example, the user may not want to get stuck in a rut of only watching teen comedies after one nostalgic viewing of *Mean Girls*, even if they do also like *Clueless*, and *Election*. By choosing a more diverse set of neighbouring user profiles on which to base recommendations, instead of just looking for recommendations among the nearest neighbours, Adamopoulos and Tuzhilin effectively mitigate both over-specalization and the popularity bias, increasing the diversity of recommendations without sacrificing prediction accuracy.

## Homogenization

Another issue for which there is some scattered evidence is homogenization. Popularity bias refers to how single items that are very popular are over-recommended. Homogenization is an effect over the dataset as a whole, where the variance of items recommended to all users combined decreases over time.

A 2008 study found that since online journals became common, which increased the availability of academic literature, citation practices have narrowed. Fewer journals, and fewer articles are being cited, suggesting that people are reading less widely, not more (Evans 2008). Evans attributes the effect to the greater efficiency of finding sources online, by following a few links, compared to browsing library stacks, where it takes longer to find specific sources, but you end up seeing a greater variety of papers in passing.

A recent study (West 2019) suggests that GoogleScholar's recommendations may have had a homogenizing effect on citation practices. More citations are going to the top 5% of papers by citation count, and a smaller proportion of papers are being cited overall since the release of GoogleScholar. When the recommendation systems we use are designed to only show us items that other users have interacted with, rather than sampling from the entire dataset equally, this narrowing of recommendations is likely to happen.

The phenomenon of "filter bubbles" or "echo chambers" is often blamed on the laziness or closed-mindedness of individuals, who can't be bothered to look beyond their social media feeds, or who don't want to do the work of consuming media that might challenge their comfortable opinions. However, when filter bubbles arise, they may result from the homogenization that is characteristic of collaborative filtering algorithms. It may not be that users fail to venture outside their bubbles, but rather that the algorithm traps users inside.

## Bias in Information Filtering

Collaborative filtering algorithms belong to the broader class of information filtering algorithms. Information filters choose items from information streams to deliver to users based on a model of the user's preferences, or a particular topic. Some common examples are a search engine returning documents that include a user provided search term, or a personalized newsfeed delivering articles on a given topic to a user's inbox. Spam filters are also information filters, but where the selected items are redirected away from users.

Information filters that continuously update their predictive model based on feedback (e.g., what the user clicks on), to improve performance during operation are alternatively called "online," "active," or "iterative". Here I will use the term iterative information filtering.

The sequence of events is a loop starting with a recommendation step based on the initial model, then the user is presented with the recommendations, and chooses some items to interact with. These interactions provide explicit or implicit feedback in the form of labels, which are used to update the model. Then the loop repeats with recommendations based on the updated model.

## Iterated Algorithmic Bias

The user's interactions change the model, based on what was recommended, which in turn affects what can be recommended at later stages. Just as in the special case of collaborative filtering, iterative information filtering introduces a selection bias (Stinson 2002; Chawla and Karakoulas 2005). Since labels are only provided for items that were recommended, the missing at random assumption is violated. This bias is investigated in Sun, Nasraoui, and Shafto (2018), who refer to it as "iterated algorithmic bias". One of the main effects of the selection bias is more homogeneous recommendations (Sun, Nasraoui, and Shafto 2018), narrowing the space of items available for recommendation.

The homogenizing bias occurs in iterative information filtering contexts generally. For some information filtering tasks, it may not be a bad thing for recommendations to become more homogenous over time. If the purpose of the filter is to find articles relevant to a very particular interest, then it might be desirable for the filter to become progressively better at picking out that one specific topic. But in contexts like GoogleScholar searches, increasingly homogenous search results for a given search term would typically be a negative outcome. For instance, if the user is doing a literature search, they want the full complement of relevant articles, not just the most cited ones. Likewise, if the user is looking for citation information for a specific article, an exact match is more desirable than the most cited match in that neighborhood.

## Statistical Bias Can Lead to Discrimination

Statistical bias has negative effects on the performance of algorithms, if uncorrected, which is bad for all users, as well as media producers and advertizers who stand to gain from accurate recommendations. The negative effects are worse for some users than others, and the implications go well beyond occasionally having to scroll past unwanted recommendations.

As algorithms mediate more and more of our access to information, access to services, and decisions about our lives, their uneven performance can become a significant equity issue. The biases described here have the greatest negative effects on users located at the margins of preference distributions: people with unusual tastes, or unique combinations of tastes. The people on the margins of distributions are literally marginalized people, whom non-discrimination law is supposed to protect (Treviranus 2014).

People from minority communities have noted that recommender algorithms do not work well for them. Noble (2018) documents the ways that search algorithms fail to serve the needs of black women. One of her examples is a hair salon owner who struggled to get her business to show up as a recommendation on Yelp when you search for "'African American,' 'Black,' 'relaxer,' 'natural,'" as keywords. Complaints about culturally inappropriate recommendations, like white hairdressers being recommended for those search terms, or Christmas movies being recommended to Jews, are common online. Popularity and homogenizing biases may be at fault in those examples. A related issue arises when the recommender system does figure out that a user belongs to a minority group, but overfits to an essentialized version of that identity. That you get recommendations for every coming age story about a gay teen after viewing a single episode of Rupaul's *Drag Race* stems from over-specialization.

There is some empirical evidence for differential effects of algorithmic bias on demographic groups. Mehrotra et al. (2017) investigate whether search engines "systematically underserve some groups of users." Ekstrand et al. (2018) find significant differences in the utility of recommendation systems for users of different demographic groups (binary gender, and age), although not exclusively benefitting the larger groups. Zafar et al. (2017) discuss "disparate mistreatment," which arises when a classifier's misclassification rates differ across social groups. An example (which stems from data bias) is how the COMPAS algorithm made more false positive errors with black defendants, labeling people who would not reoffend as being high risk, while making more false negative errors with white defendants (Angwin et al. 2016).

Perhaps the greatest source of harm is that the illusion of neutrality algorithms have is being exploited in attempts to roll back protections against discrimination. The US government has proposed changes to the Fair Housing Act that would allow for discriminatory outcomes in housing in some cases where algorithms are involved in the decisions. This includes any cases where a third party algorithm is "standard in the industry" and being used for its intended purpose. This also includes cases where a neutral third party testifies that they have analyzed the model, found that its inputs are not proxies for protected characteristics and it "is predictive of risk or another valid objective" (Department of Housing and Urban Development 2019).

The algorithms described here would pass this proposed Disparate Impact Standard test. They are in widespread use in many industries, their inputs are not necessarily proxies for protected characteristics, and they are predictive of rating accuracy, a valid objective. However as shown here, these algorithms are biased in ways that can lead to discriminatory outcomes. Clean data and standard industry practices in no way guarantee that algorithmic outcomes are equitable.

## Algorithmic Fixes

None of this is to say that there aren't algorithmic solutions to the problems caused by these biases. Contemporary collaborative filtering systems use a number of tricks for mitigating bias, including weighting items based on recency or popularity (Zhao, Niu, and Chen 2013), preferentially using items from the tail of a user's rating distribution as the basis for matching profiles (Steck 2011), or explicitly promoting new releases.

The point is that these algorithms are far from neutral. These correction techniques are needed precisely because the algorithms are biased. But these corrections can only be made when we are aware of an algorithm's biases. False claims about the neutrality of algorithms discourage further research into discovering and fixing bias in algorithms.

## Conclusions

I have discussed several types of bias inherent in the logic of a class of ML algorithm in widespread use. These biases are neither the result of biased datasets, nor of people's personal biases. Fixing biased datasets and improving the ethical behaviour of AI workers are absolutely necessary steps, but they will not eliminate all sources of bias in ML.